\documentclass[]{article}
\usepackage{graphicx}
\usepackage[latin1]{inputenc}
\usepackage{amsmath}

\usepackage{xcolor}
\usepackage[urlcolor=blue]{hyperref}      
\hypersetup{
    colorlinks = true,                    
    citecolor = {blue},
    linkcolor = {purple},
           }

\title{\bf The economics of stop-and-go epidemic
control }

\author{Claudius Gros$^*$, Daniel Gros$^\dagger$ \\
\small
$^*$Institute for Theoretical Physics, Goethe University Frankfurt, Germany \\
$^\dagger$CEPS (Centre for European Policy Studies),
Brussels, Belgium}

\begin{document}

\maketitle
\begin{abstract}
We analyse `stop-and-go' containment policies 
that produce infection cycles as periods 
of tight lockdowns are followed by periods of 
falling infection rates. The subsequent relaxation 
of containment measures allows cases to increase 
again until another lockdown is imposed and the
cycle repeats. The policies followed by several 
European countries during the Covid-19 pandemic 
seem to fit this pattern. We show that 
'stop-and-go' should lead to lower medical costs than
keeping infections at the midpoint between the highs and lows 
produced by 'stop-and-go'. Increasing the upper and reducing 
the lower limits of a stop-and-go policy by the same amount 
would lower the average medical load.  But increasing
the upper and lowering the lower limit while keeping the
geometric average constant would have the opposite 
effect. We also show that with 
economic costs proportional to containment, 
any path that brings infections back to 
the original level (technically a closed cycle) has the same
overall economic cost.
\end{abstract}

\vspace{1cm}

\newpage
\section{Introduction}

As governments grappled with the second Covid-19
wave in Europe, they usually took a far more 
gradual and graduated approach than during the 
initial phase of the pandemic. At that time 
the number of seriously ill increased so rapidly that it 
overwhelmed health systems, in particular hospital 
capacities, in several countries. The second wave, 
which started with the onset of the flu season in the autumn 
of 2020, did lead to a somewhat moderated challenge for 
health systems. 

After the first peak, the urgency to `flatten the curve'
\cite{thunstrom2020benefits} did subside to a certain 
extent. However, governments 
still felt the need to take measures to slow down the 
spread of the virus when the medical load was high. 
One key strategic issue facing 
authorities is whether they should try to preserve
a status quo, or alternate 
lockdowns with periods of easing. 

In Italy and England, the central governments instituted 
a tiered system  with different levels of social 
distancing restrictions. In regions with a higher incidence 
of Covid-19, the restrictions are tighter. Within such 
a 'traffic light' system a region (city or other 
subdivision) can graduate to a lower level of restrictions 
if its epidemiological parameters improve, and, vice versa, restrictions will be tightened if cases increase again. 
These countries thus adopted de facto a 
'stop-and-go' policy at the regional level.

A change into a higher or lower category will of 
course become more frequent the closer the 
parameters defining the various tiers are. One key 
issue for this 'regional traffic light' approach is thus 
how wide apart these parameters should be set.
We investigate this issue keeping in mind that social 
distancing measures have an economic cost, which 
increases with their severity. The choice of 
parameters should be informed by their economic cost, 
relative to the health benefits  in terms of lower 
infections, hospitalisations and deaths
\cite{gros2020strategies}.

\begin{figure*}[t]
\centerline{
\includegraphics[width=0.9\textwidth]{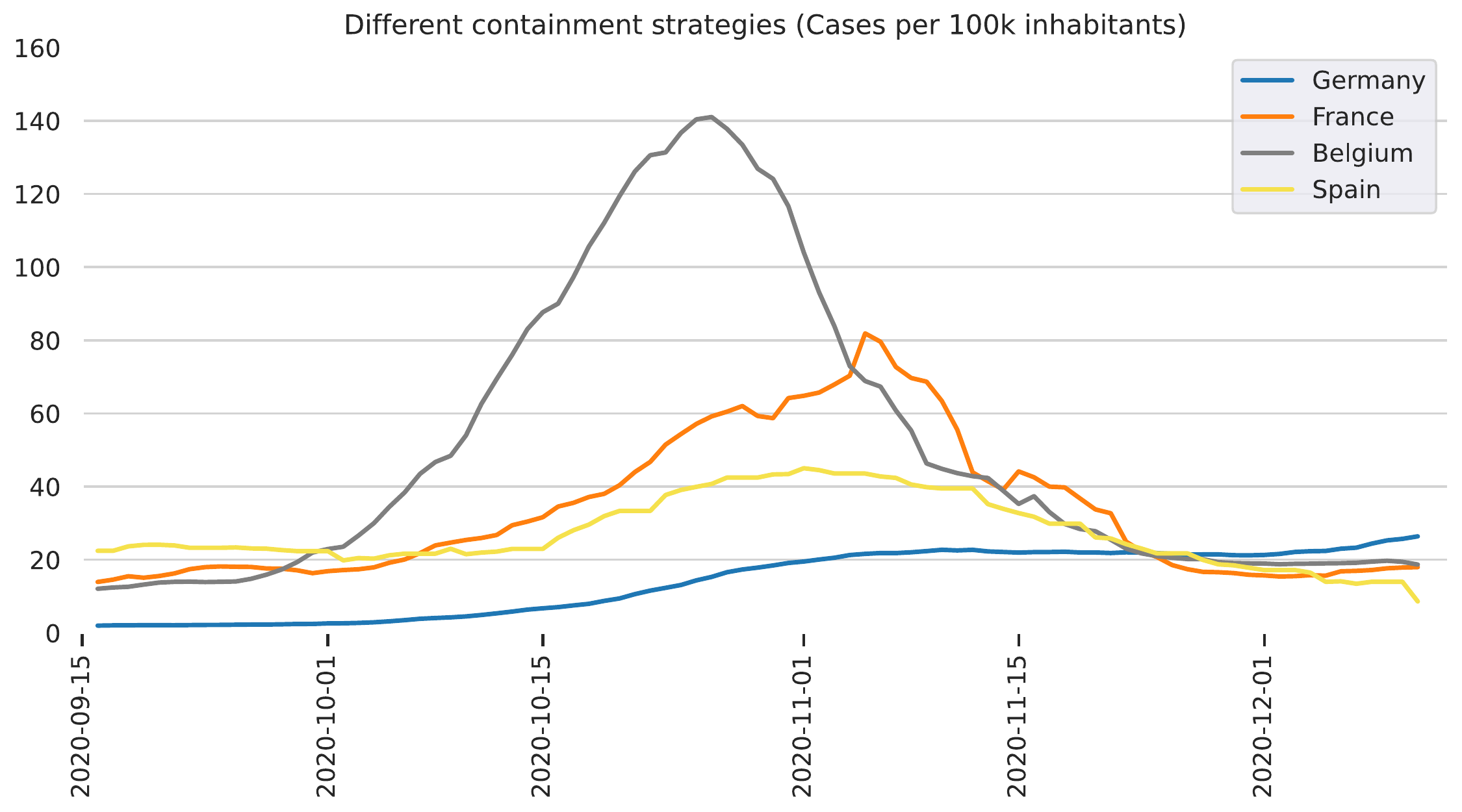}
           }
\caption{{\bf Examples of second-wave control strategies.}
For the control of the second Covid-19 wave in Europe,
most countries imposed comparatively strict lockdowns.
Illustrative examples are, as shown, France, Belgium and
Spain. As an exception, Germany, starting around the beginning of 
November 2020 opted for a `semi-lockdown',
which resulted in near constant infections.
Graphic generated using the Goethe Interactive
Covid-19 Analyser
\cite{goetheAnalyzer2020}.
}
\label{fig_danielChart}
\end{figure*}

We do not consider a general optimal control problem.
Our aim is limited to comparing policies that make 
intuitive sense and that describe the choices of 
different European countries. At the national level 
one can observe that Germany's curve remained relatively
flat compared those of France, Belgium or Spain. 
See Figure \ref{fig_danielChart}.

There is one simple economic argument that would favour 
the `stop-and-go' strategy of alternating harsh restrictions 
with broad easing. The economic cost of closing restaurants, 
closing schools or imposing restrictions on movement is the 
same whether the current rate of infections is high or low.  
This implies that one should use harsh restrictions when the 
case count is high because one would then achieve the largest 
fall in cases (in absolute numbers).

The argument against the `stop-and-go' strategy is that 
the cost of the harsh restrictions to achieve a rapid fall 
in infection is likely to be convex. A small proportional 
reduction in infections (or rather the reproduction number) 
can be achieved by measures which have little impact on 
the economy (e.g.\ mask wearing, etc.). Achieving a 
swifter deceleration in the diffusion of the virus 
requires substantially stronger restrictions of the type 
mentioned above.

One could of course argue that stop-and-go policies 
are inferior to the `East Asian' option of eradicating 
the virus \cite{shaw2020governance}, which then allows 
a total reopening of the economy. But this option had
been abandoned in Europe as the draconian measures, 
including border closures that would be required, 
have apparently been widely judged as unacceptable. Note, 
however, that it is currently yet to be settled whether
additional factors, like evolutionary adaptions, 
contribute to differences in the path the spread of
the SARS-CoV-2 pathogen took in European and East-Asian
populations \cite{yamamoto2020apparent}.

The remainder of our contribution is organised as follows:
we start by briefly reviewing the standard SIR model
to which we add a relationship that describes the economic
cost of reducing the spread of the virus. This framework
is then used to examine the economic control costs of 
two alternative policies: keeping the medical load constant \cite{budish2020maximize},
versus a stop-and-go policy. We then compute 
the medical load implied by these two policies 
over a given time path and compare the resulting 
relative economic costs against the benefits in terms 
of a lower overall number of infected. 
Finally, we consider the implications of a 
time-varying native reproduction factor,
for example an increase due to colder weather, 
leading to  more indoor interactions. 

Throughout, our purpose is not to describe and solve 
a general optimal control problem, but to compare the 
economic costs of different concrete policy options. 
Figure~\ref{fig_bangBang_ill} illustrates schematically the 'stop-and-go' epidemic control which we model below.

\section{Modelling framework}

We start with a short presentation of the standard SIR model 
where we denote with $S=S(t)$ the fraction of susceptible 
(non-affected) people, with $I=I(t)$ the fraction of the 
population that is currently ill (active cases, which are also infectious), 
and with $R=R(t)$ the fraction of recovered. Normalization
demands $S+I+R=1$ at all times. We write the continuous-time 
SIR model as
\begin{equation}
\tau\dot S = -gSI,
\qquad\quad
\tau\dot I = \big(gS -1\big)I,
\qquad\quad
\tau\dot R = I\,,
\label{SIR_model}
\end{equation}
which makes clear that $\tau$ is a characteristic
time scale. Normalization is conserved, as $\dot{S}+\dot{I}+\dot{R}=0$.
Infection and recovery rates are $g/\tau$ and $1/\tau$. 
The number of infected grows as long as $\dot{I}>0$,
namely when $gS>1$. Herd immunity is consequently
attained when the fraction of yet unaffected people 
dropped to $S=1/g$. The total number of past and present
infected is $X=1-S=I+R$.

\begin{figure*}[t]
\centerline{
\includegraphics[width=0.9\textwidth]{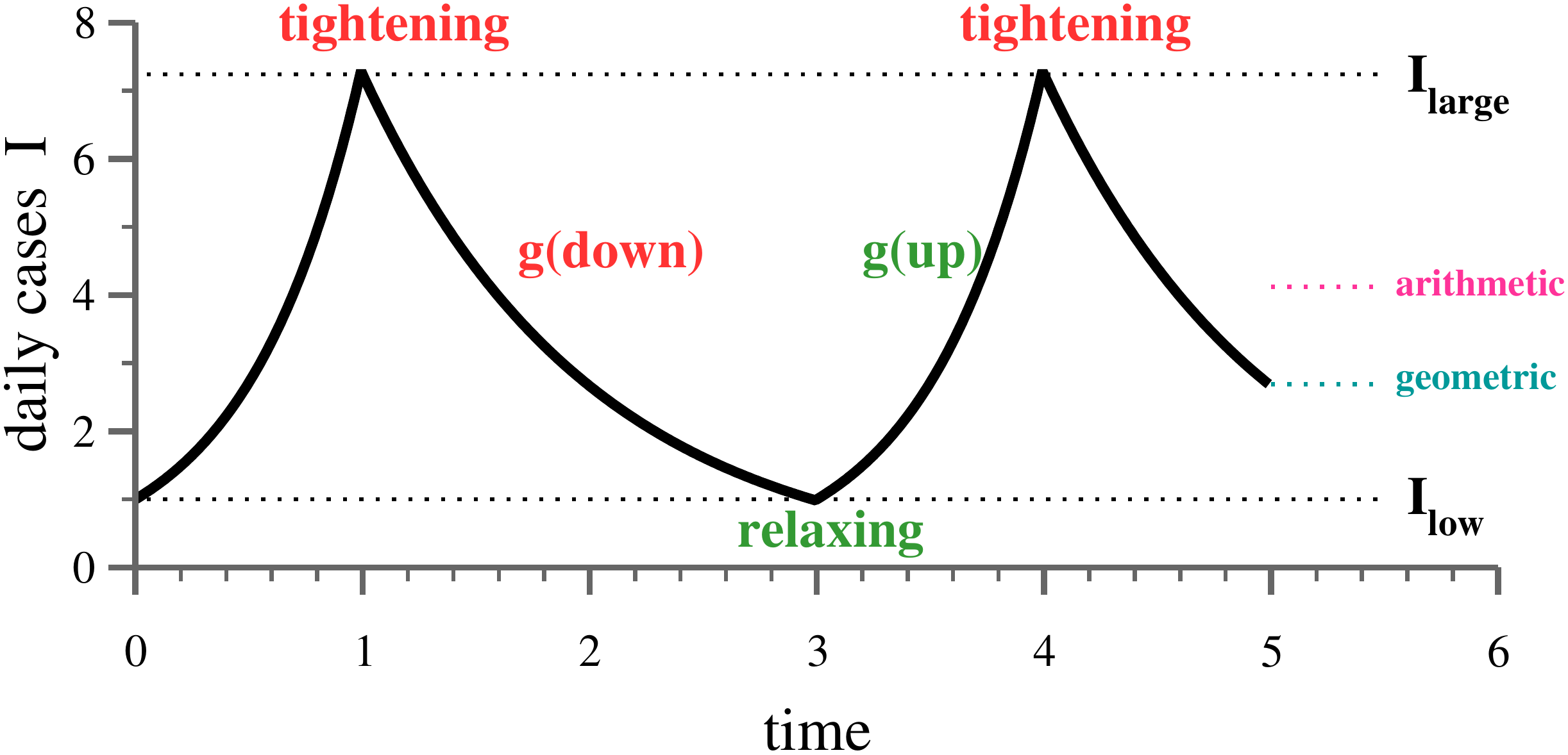}
           }
\caption{{\bf Stop-and-go epidemic control.} 
Containment policies that alternate periodically
between $I_{\rm low}$ and $I_{\rm large}$.
When relaxing, the number of daily cases $I=I(t)$
expands at a rate $g({\rm up})\equiv g_{\rm up}>1$.
Policies are tightened again when case number are 
too high. Daily case numbers then contract with
a rate $g({\rm down})\equiv g_{\rm down}<1$. In
the example shown, up- and down times are 
$T_{\rm up}=1$ and $T_{\rm down}=2$. For a
presumed time scale of one month the period
$T_{\rm up}+T_{\rm down}$ of the control cycle
would be here three months. Also indicated are
the arithmetic and the geometric means of 
$I_{\rm low}$ and $I_{\rm large}$, as used in
Sect.~\ref{sect_band_I}. Note that contant control
is recovered in the limit $I_{\rm low}\to I_{\rm large}$.
}
\label{fig_bangBang_ill}
\end{figure*}

From (\ref{SIR_model}) one sees that $g/\tau$ governs 
the transition between two compartments, from susceptible 
to infected. This transition rate is constant within
the basic version of the SIR model. There are two venues
to relax this condition:
\begin{itemize}
\item {\bf Non-linear reproduction rates.} The basic
reproduction factor $g$ may depend functionally on
the actual number of infected $I$
\cite{capasso1978generalization,hethcote1991some},
or on the total number $X$ \cite{gros2020strategies}.
This happens when societies react on an epidemic
outbreak.
\item {\bf Time-dependent reproduction rates.} 
From the viewpoint of the pathogen, certain changes 
in the transmission rate $g=g(t)$ are external, 
e.g.\ because hosts decide more often to quarantine.
\end{itemize}
Here we focus on time-dependent $g=g(t)$, mostly
as induced by stop-and-go politics, as illustrated
in Figure~\ref{fig_bangBang_ill}. We assume a
stop-and-go cycle which repeats after
$T=T_{\rm up}+T_{\rm down}$:
\begin{equation}
g(t) = \left\{\begin{array}{rcl}
g_{\rm up}>1   && t\in[0,T_{\rm up}] \\[0.5ex]
g_{\rm down}<1 && t\in[T_{\rm up},T] 
\end{array}\right.\,.
\label{g_bangBang}
\end{equation}
The time spans during which epidemics 
expands/contracts are respectively 
$T_{\rm up}$ and $T_{\rm down}$. The
policy is cyclic if $I(0)=I(T)$, viz
when the starting case number is reached
again.

\subsection{Low incidence approximation}

We assume that infection counts are substantially
lower than the population, viz that $I\ll1$.
For example, even in a highly affected country like Italy, 
the total number of daily infections has rarely exceeded 
thirty thousand \cite{github2020}, which corresponds to
less than 0.0005 of total population. The total number of 
cases has reached 1.5 million, which is equivalent to 
$S\approx0.975$. Given that the economic costs associated
with raising infection numbers are based on estimates,
as discussed in the subsequent section, the low
incidence approximation provides in comparison
sufficient accuracy.

Within the low-incidence approximation the number of 
susceptibles remains constant and we
can set, without loss of generality $S\to1$. 
Hence we need to deal only with
\begin{equation}
\tau\dot I = \big(g -1\big)I,
\qquad\quad
I(t) = I(t_0)\,\mathrm{e}^{(t-t_0)(g-1)/\tau}\,.
\label{I_model}
\end{equation}
In the quasi-stationary state we have simple 
exponential growth/decay. For stop-and-go control
the reproduction factor is piece-wise constant, which
allows us to evaluate explicitly the time evolution of
case numbers, and with this the associated economic
costs.

\section{Economic costs of disease control}

The economic costs of imposing social distancing 
on a wider population, closing restaurants or retail 
trade, are at the core of policy discussions. The key 
issue here is how these costs vary with the social 
distancing measures (so-called Non-Pharmacological Interventions, 
NPI) imposed. They increase in severity from mask requirements, 
abstaining from travel or restaurant meals to more invasive 
interventions like closure of schools, lockdowns or curfews.
Limiting social interaction necessarily reduces economic activity.
This suggests that the economic cost of the social distancing, 
measured as the proportional loss of GDP, should increase with 
the reduction in the transmission rate described by $g$,
\begin{equation}
E = c_e\,\frac{g_0-g}{g_0}\,,
\label{economicCosts}
\end{equation}
where $g_0$ is the native reproduction factor. 
Here we assumed that social distancing costs are 
proportional to the percentage-wise reduction
of the reproduction factor, viz to $(g_0-g)/g_0$.

Our basic assumption, that social distancing costs are 
proportional to the reduction in the reproduction 
parameter, differs from the assumptions underlying matching 
models, as used in
\cite{gollier2020pandemic} or \cite{acemoglu2020multi},
which typically arrive at a quadratic relationship 
between the economic cost and the reduction in contagion, whereas \cite{rowthorn2020cost} postulates simply a convex cost curve.

A concrete example can illustrate the key mechanism 
behind the matching framework, which usually assumes that
 the 'lockdown' takes the form of the confinement 
 of a proportion of the population, 
and that contagion is possible only outside.  
If one half of the population has to stay at home, 
only the other half can go out and get potentially infected.  
But the half which is not confined will find only one half of 
their potential partners outside, 
resulting in one fourth of the number of matches.
Contagion should thus be reduced by a factor of four when
confining one half of the population.

In the matching framework, the economic cost is assumed to be 
proportional to the percentage of the population which is 
confined - not the number of matches.  
This framework thus separates social activity (matches, meetings of people) 
from economic activity, assuming that contagion 
is fostered only through social activity.
Another implicit assumption behind this view is that those 
who are not confined will not have longer meetings with the 
ones they still find outside; 
or that those who are free to move accept to have only 
half of matches, and do not decide to meet somebody else 
if their preferred match is not available. 
These implicit assumptions are crucial.  
For example, contagion would only be halved if those 
not confined would meet twice as long with the remaining
matches they find.
Some authors, e.g.\ \cite{acemoglu2020multi}
acknowledge these considerations by allowing 
for different economies of scale in matching.

Our view of lockdown or rather social distancing is that 
governments mandate the closure of some part of the economy, 
in reality  mainly the services sector (restaurants, bars, shops, etc.). 
This is different from a strict confinement of a part of the population.  
A restaurant which is closed (or limited in its opening hours) 
results in less value added created 
and diminishes at the same time the potential for contagion. 
But the restaurant owners and their workers are not confined, 
they can meet others. 
There is thus no quadratic effect in terms of contacts.
We thus start from the assumption that economic activity 
involves occasions for contagion, 
implying that the loss of economic activity 
should be directly proportional to the reduction in 
occasions for contagion and thus the effective reproductions rate.
For related approaches see 
\cite{gros2020strategies,strong2020estimation,krueger2020macroeconomic,Eichenbaum2020}.

This view of 'lockdown' corresponds closer to the measures 
adopted by many governments during this second wave.  
The matching model might have been more appropriate 
during the first wave when indeed in some countries 
large parts of the entire population were forbidden 
to leave their home, except for essential business.

Finally we note that social distancing measures (NPIs) 
cannot affect the number of infected, 
only the rate at which their number grows over time. 
Eq.~\ref{economicCosts} implies that the only way 
to avoid all contagion is to completely shut down 
the economy. The parameter $c_e$ represents a scaling 
factor, which depends on the structure of the 
economy (importance of services requiring close contact, 
like tourism) and the degree to which the population 
effectively adheres to official restrictions.
It hast been estimated that $c_e$ is of the order of
$0.25$ \cite{gros2020strategies}.

\subsection{Uniform control}

We first briefly examine the implications of a policy which
keeps the number of infected  
\cite{budish2020maximize} and thus the medical load constant.
Such a policy is of course not optimal, but it serves as a useful
benchmark for our more general results. 
It can be considered as the limiting case 
of the 'traffic light' system
in which the difference in parameters between tiers or 
levels becomes vanishing small.

Formally, uniform control implies a constant 
fraction $I\to I_{\rm const}$ of infected. This is 
achieved for $g=1$, independent of the value of 
$I_{\rm const}$. The economic cost $E_{\rm const}$ 
per time unit is therefore
\begin{equation}
E_{\rm const} = c_e\,\frac{g_0-g}{g_0}\Big|_{g=1}
\ =\  c_e\,\frac{g_0-1}{g_0}\,,
\label{E_const}
\end{equation}
Note that $E_{\rm const}$ is independent of the
value of the medical load one wants to retain.
As already mentioned above, the economic costs 
of keeping the reproduction factor at one is 
independent of how many infected there are.

\subsection{stop-and-go control}

stop-and-go, or 'bang bang', control corresponds to an
on-off policy as illustrated in Figure~\ref{fig_bangBang_ill} above, 
which can be described within the  framework developed here
by the following rule:
Control is increased when $I=I(t)$ reaches an 
upper threshold $I_{\rm large}$, and decreased
when $I=I(t)$ falls below a lower threshold
$I_{\rm low}$. 

We denote with $g_{\rm down}\!<\!1$ the small 
reproduction rate corresponding to strong
control, and with $g_{\rm up}\!>\!1$ the 
large reproduction factor corresponding to 
weak control. Using the low-incidence approximation 
(\ref{I_model}) we find 
\begin{equation}
T_{\rm up} = \frac{\tau}{g_{\rm up}-1}\ln\left(
\frac{I_{\rm large}}{I_{\rm low}} \right),
\qquad\quad
T_{\rm down} = \frac{\tau}{g_{\rm down}-1}\ln\left(
\frac{I_{\rm low}}{I_{\rm large}} \right)
\label{T_up_down}
\end{equation}
by integrating $I(t)$
from $t=0$ to $t=T_{\rm up}$, the time needed for 
$I(t)$ to grow from $I_{\rm low}$ to $I_{\rm large}$,
with a respective expression for the down-time 
$T_{\rm down}$. 

As one would expect from an exponentially growing 
variable, the time needed to evolve from one value to 
another is a function of the logarithm of the ratio 
of end- and starting points, being inverse to the effective
growth rate of infections (held constant by restrictions).
Given that the economic costs are constant per unit of time (lower during the period of allowing infections to increase, higher during the restrictive phase)  one arrives at the following solution for the total economic costs: 

\begin{equation}
E_{\rm bang} = \frac{c_e}{g_0}\Big[
(g_0-g_{\rm up}) T_{\rm up} +
(g_0-g_{\rm down}) T_{\rm down}
\Big]\frac{1}{T_{\rm up}+T_{\rm down}}
\label{E_bang_bang}
\end{equation}
when considering an entire period or cycle, up and down when  stop-and-go control is applied. 

\subsection{Vanishing cost differential}

The cost difference between bang-bang and
constant control is 
\begin{equation}
E_{\rm bang} - E_{\rm const} = 
\frac{c_e}{g_0}\left[1-
\frac{g_{\rm up} T_{\rm up} + g_{\rm down} T_{\rm down}}
{T_{\rm up}+T_{\rm down}} \right]\,.
\label{E_band-const}
\end{equation}
Note that
$\ln(I_{\rm large}/I_{\rm low})=
-\ln(I_{\rm low}/I_{\rm large})$,
which implies that both $\tau$ and
$\ln(I_{\rm large}/I_{\rm low})$
drop out of (\ref{E_band-const}), which 
vanishes as
\begin{equation}
E_{\rm bang} - E_{\rm const} = 
\frac{c_e}{g_0}\left[1-
\frac{g_{\rm up  }(g_{\rm down}-1) 
    - g_{\rm down}(g_{\rm up  }-1)}
{g_{\rm down}-g_{\rm up}} \right]\equiv 0\,.
\label{E_band-const_0}
\end{equation}
This implies that both control types, constant and
stop-and-go control, come with the same economic costs. 

\subsection{Neutrality theorem\label{sect_theorem}}

The result, that the cost differential between
constant and stop-and-go control vanishes, can 
be generalised if we rewrite (\ref{I_model}) as
\begin{equation}
\tau\dot I_{\rm log} = g -1,
\qquad\quad
I_{\rm log} = \ln(I)\,,
\label{I_log}
\end{equation}
where $g=g(t)$ is now an arbitrary function
of time. We then have
\begin{equation}
I_{\rm log}(t_{\rm end})- I_{\rm log}(t_{\rm start}) =
\int_{t_{\rm start}}^{t_{\rm end}}
\dot I_{\rm log} dt =
\int_{t_{\rm start}}^{t_{\rm end}}\frac{g(t)-1}{\tau} dt\,,
\label{I_log_start_end}
\end{equation}
which proves that 
\begin{equation}
\frac{1}{t_{\rm start}-t_{\rm end}} 
\int_{t_{\rm start}}^{t_{\rm end}} g(t)dt = 1
\label{g_start_end}
\end{equation}
for closed trajectories, viz when
$I_{\rm log}(t_{\rm end})= I_{\rm log}(t_{\rm start})$.
Comparing with (\ref{E_const}) shows that
average economic costs are independent of
which timeline $g(t)$ is used for controlling the
epidemic.  
Note that the case of constant control, 
$g(t)\equiv g$, is included as a special case.
If the economic costs of control are proportional to the reduction
in the reproductions rate, one finds thus a 'neutrality theorem' for epidemic control.  All trajectories which return to the point of departure (in terms of 
the infection rate) will lead to the same economic cost.

\section{Mean number of infected under different control policies}

The number of infected becomes the key criterion if 
the economic cost of different control policies 
(over complete cycles) is the same.

For constant infection rates $g$, the cumulative number of 
infected between two times $t=t_0$ and $t=t_1$ is
\begin{eqnarray}
\nonumber
X_{0,1} &=&\int_{t_0}^{t_1} I(t)dt = 
I(t_0)\, \int_{t_0}^{t_1} \mathrm{e}^{(t-t_0)(g-1)/\tau}\,dt
\\[0.5ex]
&=& \frac{I_0\tau}{g-1} \left(
\mathrm{e}^{(t_1-t_0)(g-1)/\tau}-1\right)
= \frac{(I_1-I_0)\tau}{g-1} 
\label{X_01}
\end{eqnarray}
when using (\ref{I_model}) and that 
$I_1= I_0\exp((t_1-t_0)(g-1)/\tau)$.
Noting that (\ref{T_up_down}) holds generally for
constant $g$, we have
\begin{equation}
\Delta T_{0,1} = t_1-t_0 = \frac{\tau}{g-1}
\ln\left(\frac{I_1}{I_0} \right)\,,
\label{Delta_T}
\end{equation}
which leads to
\begin{equation}
\frac{X_{0,1}}{\Delta T_{0,1}} = 
\frac{I_1-I_0}{\ln(I_1/I_0)}
\label{X_01_Delta_t}
\end{equation}
for the overall number of infected,
on the average per time unit. Note that
\begin{equation}
\ln\left(\frac{I_1}{I_0} \right) = 
\ln\left(\frac{I_1-I_0+I_0}{I_0} \right)
\approx \frac{I_1-I_0}{I_0} 
\label{log_approx}
\end{equation}
for $I_1\approx I_0$, from which the limit
\begin{equation}
\lim_{I_0\to I_1} \frac{X_{0,1}}{\Delta T_{0,1}} = I_0
\label{}
\end{equation}
is recovered.
Empirically it been observed that the
cumulative medical load during the 'down phase'
is about 30\% higher than the cumulative load which follows the peak
\cite{gros2020predicting}.

\subsection{Bang-bang infection numbers\label{sect_band_I}}

For stop-and-go control we have two periods
with constant $g$, when $I$ goes up and 
respectively down. With (\ref{X_01_Delta_t}) 
we find that the total cumulative fraction of infected is determined by;
\begin{equation}
X_{\rm bang} = 
\frac{I_{\rm large}-I_{\rm low}}{\ln(I_{\rm large}/I_{\rm low})}
T_{\rm up} +
\frac{I_{\rm low}-I_{\rm large}}{\ln(I_{\rm low}/I_{\rm large})}
T_{\rm down}\,
\label{X_bang}
\end{equation}
for bang-bang control, and hence
\begin{equation}
\bar{I}_{\rm bang} =
\frac{X_{\rm bang}}{T_{\rm up}+T_{\rm down}} =
\frac{I_{\rm large}-I_{\rm low}}{\ln(I_{\rm large}/I_{\rm low})}
\label{X_bang_Delta_t}
\end{equation}
for the time-averaged number $\bar{I}_{\rm bang}$ of
infections. 

Here we are interested in on-the-average
stationary control strategies, for which
the level $I$ of infections is kept at the
average. In Sect.~\ref{sect_theorem} we did
show that stationary control results always
in identical economic costs. This holds
however not for the medical load,
which can be considered to be the proportional
to the average infection number $\bar{I}$.
It is in particular of interest to compare the
medical $\bar{I}_{\rm bang}$, of stop-and-go
control, with the policies keeping $I$ at
a constant, intermediate level or benchmark. 
A natural choice for this benchmark could the 
arithmetic mean, or midpoint, $I_{\rm mid}=(I_{\rm large}+I_{\rm low})/2$.  
However, we will consider as well the geometric mean
$I_{\rm geo}=\sqrt{I_{\rm large}I_{\rm low}}$.

For the arithmetic mean, the ratio
$\bar{I}_{\rm bang}/\bar{I}_{\rm mid}$
in mean infection numbers is
\begin{eqnarray}
\nonumber
\delta\bar{I}\,=\,
\bar{I}_{\rm bang}\big/\bar{I}_{\rm mid}&=&
\frac{X_{\rm bang}}{T_{\rm up}+T_{\rm down}} 
\,\frac{2}{I_{\rm large}+I_{\rm low}}
\\[0.5ex] &=&
\frac{x-1}{\ln(x)}\,\frac{2}{x+1}\,\le\,1\,,
\label{Delta_I_mid}
\end{eqnarray}
when denoting $x=I_{\rm large}/I_{\rm low}$.
In the limes $I_{\rm large}\to I_{\rm low}$, viz
$x\to1$, one has $\lim_{x\to1} \delta\bar{I}=1$, as
expected. The limes $x\to1$ is performed using the 
small $x-1$ expansion $\ln(x)=\ln(1+(x-1))\approx x-1$.

\begin{figure*}[t]
\centerline{
\includegraphics[width=0.9\textwidth]{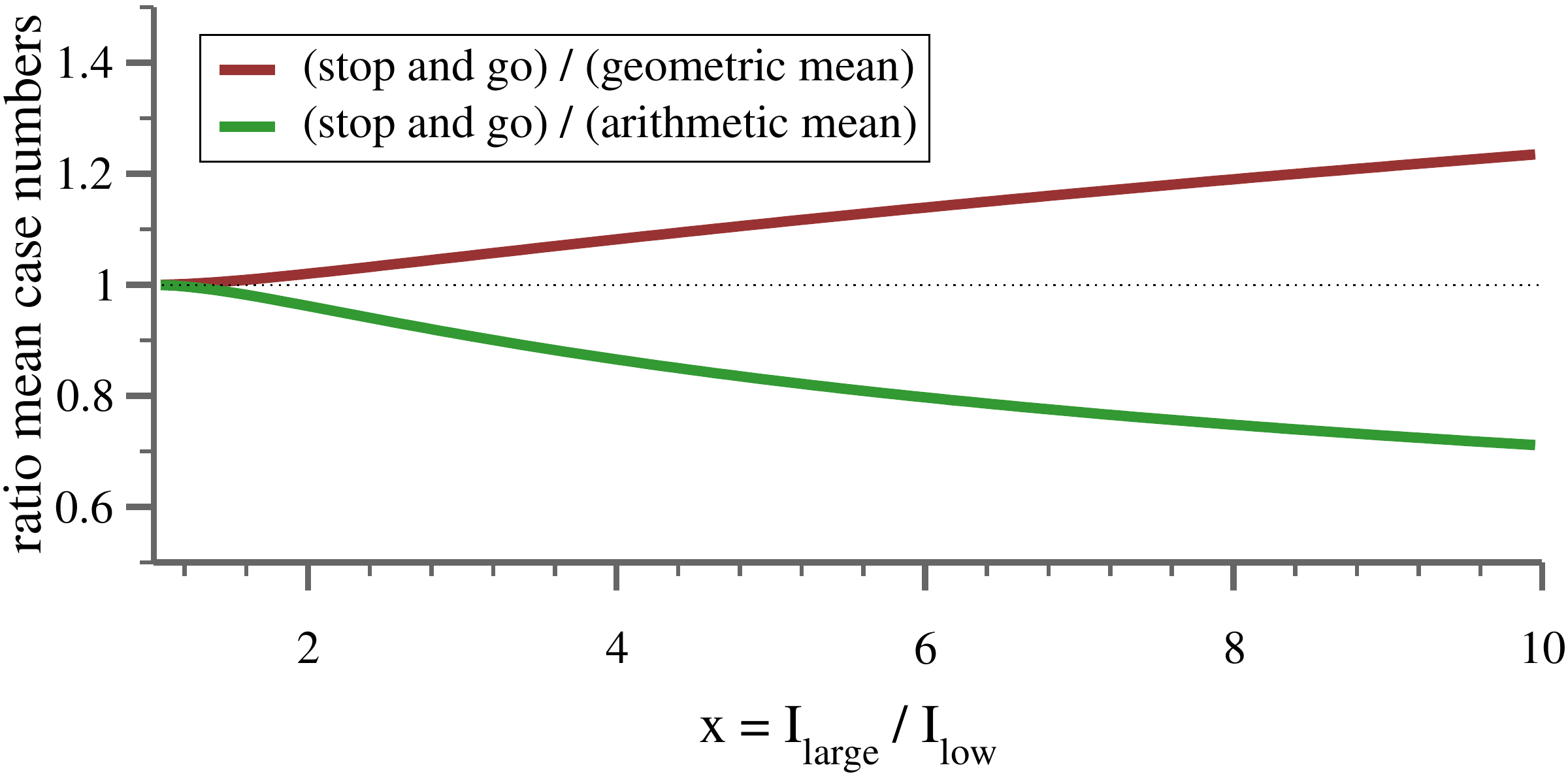}
           }
\caption{{\bf Relative case numbers for distinct control policies.} 
For stop-and-go control case numbers oscillate between
$I_{\rm low}$ and $I_{\rm large}$. Alternatively considered
are constant incidences, either at the arithmetic
mean, $(I_{\rm low}+I_{\rm large})/2$, or at the geometric
mean, $\sqrt{I_{\rm low}I_{\rm large}}$. Shown are the
ratios of the respective per time case numbers, as
given by Eqs.~(\ref{Delta_I_mid}) and
(\ref{Delta_I_geo}).
}
\label{fig_ratio_I}
\end{figure*}

That $\bar{I}_{\rm bang}/\bar{I}_{\rm mid}$ is 
strictly smaller than unity for $x>1$ can be seen, 
e.g., by plotting (\ref{Delta_I_mid}) as a function 
of $x$, as done in Figure~\ref{fig_ratio_I}. Numerically
one finds a reduction of 29\% at $x=10$, which
is somewhat higher than ratio of 7 to 1 observed 
in the actual values displayed in Fig.~\ref{fig_danielChart}.
Note that one
has $2/(x+1)\le 1$ when $x\ge1$, 
for the second term in (\ref{Delta_I_mid}), with 
$(x-1)/\ln(x)\le1$ holding for the first term. The
last relation holds because the log-function is concave,
with a slope $d\ln(x)/dx=1$ at $x=1$.

The result that $\bar{I}_{\rm bang}\le \bar{I}_{\rm mid}$ 
implies that a policy of stop-and-go is superior 
to (less bad than) a policy 
of keeping infections constant half-way between the peak 
and trough. The two policies would have the same economic 
cost, but stop-and-go would lead to a lower overall 
medical load. Neither policy would of course be optimal 
in an unconstrained policy space.  But our aim is merely
to consider policies that have been adopted.
The intuition behind this result can be seen from
Figure~\ref{fig_bangBang_ill} above.  
The infection curve lies more
time below than above the arithmetic average, a
well-known property of exponential growth.

The constant control policy serves only as benchmark.
The results are much more general: 
As can be seen from (\ref{Delta_I_mid}) 
the average medical load falls as the ratio 
of the upper limit to the lower limit increases, while holding
the (arithmetic) average constant. This implies that 
the medical load resulting from
a 'stop-and-go' policy improves  as one increases the 
upper and reduces the lower limit by the same amount.

Regarding the comparison to the
geometric mean $I_{\rm geo}$, we consider
\begin{eqnarray}
\nonumber
\bar{I}_{\rm bang}\big/\bar{I}_{\rm geo} &=&
\frac{I_{\rm large}-I_{\rm low}}{\ln(I_{\rm large}/I_{\rm low})}
\,\frac{1}{\sqrt{I_{\rm large}I_{\rm low}}}
\\[0.5ex] &=&
\frac{x-1}{\ln(x)}\,\frac{1}{\sqrt{x}}\,\ge1\,,
\label{Delta_I_geo}
\end{eqnarray}
where we used $x=I_{\rm large}/I_{\rm low}$,
as for (\ref{Delta_I_mid}). The functional dependence
is included in Figure~\ref{fig_ratio_I}. It can be
seen that it is favorable to keep the incidence at
the geometric mean, instead of letting it oscillate 
between $I_{\rm low}$ and $I_{\rm large}$. Moreover, 
the ratio of the respective average medical loads increases
only modestly as the ratio between upper and lower limit 
increases. Letting case number vary by a factor of ten
leads to an increase of 24\%.  It should not be surprising 
that keeping infections at the geometric mean is worse, given that
the distance between the two means increases as the ratio of the 
upper to the lower limit becomes larger. For  $9=I_{\rm large}/I_{\rm low}$,
the arithmetic mean is equal to 5 whereas the geometric mean is 3. Choosing the geometric as the intermediate point thus amounts 
to choosing a higher benchmark. 

From a general viewpoint, if follows from (\ref{Delta_I_geo}) that
the average medical load increases as the ratio 
of the upper limit to the lower limit increases, when holding
the geometric average constant. This implies that 
the medical load resulting from a 'stop-and-go' policy 
increases as one increases the upper and 
reduces the 
lower limit by the identical factor.
These considerations suggest that the decision regarding how wide apart 
to set the limits of a 'stop-and-go' or a regional 'traffic light'
policy depends on whether one wants to keep the arithmetic or the
geometric average constant.  

\subsection{Significance}

The results discussed above apply only within the 
limits of hospital capacity. Once that 
limit has been reached any further increase in the 
medical load would imply rapidly rising medical and 
ethical costs.  A key assumption made above is that the 
medical load is proportional to the number of infections.  
This is a reasonable assumption as long as each infected 
can be cared for, even when severe symptoms requiring 
hospitalisation develop. Hospital capacity and especially 
the number of intensive care units (ICU) create thus 
a practical upper bound on political choices. The 
upper (and lower) limit cannot be increased beyond the 
range given by hospital capacity. The question we address 
is to what extent this range should be used by the authorities.

A lower medical load constitutes a sufficient criterion 
for preferring the stop-and-go  policy, given that the 
respective economic costs vanish for different control
paths as long as the initial infection incidence is reached again.
Note that the medical 
load can be translated into economic costs
\cite{Eichenbaum2020,atkeson2020will,Brodeur,baldwin2020keeping}.

Most of the literature focuses on the economic 
value of the lives lost. However, that might be a 
mistake \cite{gros2020strategies}, as infections 
with less severe symptoms can also lead to considerable economic costs. One example of the economic cost of 
infections would be the loss of working time of 
those infected and with mild symptoms, when these
individual have to self-isolate and
cannot work for a certain time. This loss could be calculated as  
the number of weeks of working time due to symptoms 
(and self-isolation needs) and would be equal to a 
proportion of GDP \cite{gros2020strategies}. To this one would have to add the hospitalisation costs for 
those with stronger symptoms and finally the economic 
value of lives lost. The lower medical load implied by 
a stop-and-go policy would thus also lead to lower overall economic costs.

A key difference between the present study and most existing 
literature concerns the kind of policies examined. Here
we focus on a comparison between two representative, real-world
policies, stop-and-go vs.\ constant control. 
Optimal control, 
which usually does not result in reversals of restrictions,
is in contrast the goal for the majority of studies published
so far. Contributions like \cite{gollier2020pandemic} and
\cite{acemoglu2020multi} employ the matching framework,
which is based on confinement as the main policy instrument 
and allows for economies of scale in lowering contagion as 
explained above. In such a framework it is clear that
the optimal policy would be to impose tight 
restrictions from the start until the desired incidence 
is attained. For example, 
\cite{gollier2020pandemic} 
finds that ``the optimal 
confinement policy is to impose a constant rate of lockdown 
until the suppression of the virus in the population.''.

Other contributions, which do not incorporate economies of 
scale in containment policies (either because they use constant 
returns in matching or because of a different view of social
distancing restrictions) arrive at somewhat differentiated conclusions. 
For example, \cite{Eichenbaum2020} where containment
works like a consumption tax, find that containment should
build up gradually and peak early when there
exists the perspective of a vaccine being discovered.

\section{Changing epidemic parameters}

The native reproduction factor $g_0$ is,
as a matter of principle, an intrinsic 
property of the virus. As such it can be
measured only at the very start of a pandemic, 
namely when nobody is yet aware of what is 
happening. However, this reproduction factor
can change over time, for example with the season.
It is well known that the danger of contagion is 
much higher indoors than outdoors. It is in fact
nearly impossible to catch the Coronavirus
outside \cite{coronaOutside}. 
In the case of influenza virus this change in the reproduction 
rate leads to the typical 'flu season', 
which starts with the onset
of colder weather (late in the year in the Northern Hemisphere \cite{sajadi2020temperature}). 
For the case of the Covid-19 virus other seasonal factors have also been mentioned, for example fluctuations in UV light \cite{Holmstrom2020}, \cite{karapiperis2020assessment}, for a survey see \cite{carlson2020misconceptions}.

One thus needs to consider time-dependent $g_0$,
which would correspond to the underlying spreading rate
in a given societal state, i.e without 
spontaneous behavioural modifications or explicit,
government-imposed restrictions. 

We assume that $g_0$ changes right at the top for
the case of bang-bang control. Going up/down we 
then 
have $g_0=g_0^{\rm(up)}$ and $g_0=g_0^{\rm(down)}$.
As a further simplification we postulate 
\begin{equation}
T_{\rm up} = T_{\rm down}, \qquad\quad
g_{\rm up}-1 = 1-g_{\rm down}\,,
\label{T_up_equals_down}
\end{equation}
which is equivalent to $g_{\rm up} + g_{\rm down} = 2$.
Compare (\ref{T_up_down}).

\subsection{Costs for the economy}

Given that we assume that $T_{\rm up} = T_{\rm down}$,
the per time economic costs of keeping constant
infection numbers is
\begin{equation}
E_{\rm const} = \frac{c_e}{2}\left[
\frac{g_0^{\rm(up)}-1}{g_0^{\rm(up)}} +
\frac{g_0^{\rm(down)}-1}{g_0^{\rm(down)}}
\right]\,.
\label{E_const_pp}
\end{equation}
For bang-bang control we have instead
\begin{equation}
E_{\rm bang} = \frac{c_e}{2}\left[
\frac{g_0^{\rm(up)}-g_{\rm up}}{g_0^{\rm(up)}} +
\frac{g_0^{\rm(down)}-g_{\rm down}}{g_0^{\rm(down)}}
\right]\,.
\label{E_bang_pp}
\end{equation}
The difference is
\begin{equation}
E_{\rm bang} - E_{\rm const} = \frac{c_e}{2}\left[
\frac{1-g_{\rm up}}{g_0^{\rm(up)}} +
\frac{1-g_{\rm down}}{g_0^{\rm(down)}}
\right]\,,
\label{E_bang-const_pp_0}
\end{equation}
which simplifies to
\begin{equation}
E_{\rm bang} - E_{\rm const} = \frac{c_e}{2}\left[
\frac{1}{g_0^{\rm(down)}} -
\frac{1}{g_0^{\rm(up)}}
\right]\, 
\underbrace{\big(1-g_{\rm down}\big)}_{>\,0}
\label{E_bang-const_pp_1}
\end{equation}
when using the $T_{\rm up} = T_{\rm down}$
condition that $g_{\rm up}-1 = 1-g_{\rm down}$,
see (\ref{T_up_equals_down}).
Going doing down we restrict, viz 
$g_{\rm down}<1$ holds.

Bang-bang control is therefore favorable when 
$g_0^{\rm(down)}> g_0^{\rm(up)}$. 
This result would support the pattern observed in Europe where during the summer (when $g_0$ was lower than before) governments eased restrictions, imposing them again during the fall when $g_0$ increased again. The economic cost of the restrictions could thus be concentrated during the period when they had a higher yield in terms of infections avoided.

\section{Conclusions}

During the first wave of the Covid-19 pandemic the key 
concern was to 'flatten the curve'.  Harsh lockdown 
measures were needed when health systems were
overwhelmed by the sudden increase in hospitalisations, 
many of which required intensive care units. The second 
wave resulted in a somewhat lower medical load, but 
it proved nevertheless indispensable to re-introduce some 
social distancing measures as otherwise the case load 
would have continued to increase at a near exponential rate. 
Countries have taken different approaches in this regard.
In some, the measures have been just enough to stabilise 
infections. In others, the measures led to a strong 
fall in new cases and governments lifted
restrictions - which in some cases led to a renewed 
increase. 

Several countries introduced, interestingly, regionally 
graduated systems, which allow regions and 
cities to 
oscillate between periods of harsh restrictions
that are triggered when infection numbers exceed certain 
thresholds, and periods of lower restrictions, which start
when numbers have fallen again, now below a given threshold.
These kind of quasi-automatised threshold containment policies
are a graded realisation of the stop-and-go
policy examined in the present study. 

In the context of our model we have shown 
that any time path of restrictions which 
returns to the point of departure in terms 
of the infection rate, a scenario likely to 
occur within rule-based 'traffic light' 
systems, implies the same overall economic 
costs. Furthermore, our analysis suggests 
that stop-and-go policies might not be as 
costly as it could appear at first sight, at 
least if compared to the alternative of keeping the 
incidence rate constantly at the midpoint. 
The economic cost would be the same, but 
the overall medical load of stop-and-go should be lower.
The result is reversed for the alternative of
keeping
infections at the (lower) geometric mean. Compare
Figure~\ref{fig_bangBang_ill}.

Applied to regional 'traffic light' systems, our 
results imply that increasing the upper, and
reducing at the same time the lower threshold 
by the identical absolute amounts $\Delta I$, via 
$I_{\rm upper}\to I_{\rm upper}+\Delta I$
and
$I_{\rm lower}\to I_{\rm lower}-\Delta I$,
leads to a lower medical load. The opposite 
would be true if one were to increase the 
upper, and at the same time reduce the 
lower threshold by the same proportion, $f_I>1$,
this time using the rescaling
$I_{\rm upper}\to I_{\rm upper}f_I$ and
$I_{\rm lower}\to I_{\rm lower}/f_I$.
The latter procedure would increase
the arithmetic average since the absolute 
increase in the upper limit would be now higher 
than the fall in the lower limit. This implies 
that any choice regarding the range
between the upper
and lower thresholds in a regional 'traffic light' 
must take this difference between arithmetic and geometric
mean into account. In political reality the arithmetic average
seems to be the more important benchmark as many comparisons across
countries focus on the average number of infections over a given period.

We have purposely not formulated an optimal control 
problem because our aim was to provide a framework 
for studying policies that have been widely adopted 
in the real world. However, the results concerning the
arithmetic mean could also be interpreted as implying 
that the `optimal control' (always within the class of 
policies we consider), would be to increase the range 
(viz $\Delta I$) up to the maximum permitted by hospital 
capacity.  

We also show that it makes sense for policies to react to 
seasonal variations in the native reproduction factor.
The economic cost of tight social distancing should be incurred 
in winter, when infections would otherwise be high and rising.

\section*{Acknowledgements}
Funding received from the European Union's Horizon 2020 
research and innovation program, PERISCOPE: 
Pan European Response to the Impacts of Covid-19 
and future Pandemics and Epidemics, under 
grant agreement No. 101016233, 
H2020-SC1-PHE CORONAVIRUS-2020-2-RTD.
We thank Charles Wyplosz and an anonymous 
reviewer for valuable comments. 
This research was motivated by a discussion 
of the D.G.\ with Jakob von Weizsaecker.


\bibliographystyle{unsrt}

\begin{thebibliography}{10}

\bibitem{thunstrom2020benefits}
Linda Thunstr{\"o}m, Stephen~C Newbold, David Finnoff, Madison Ashworth, and
  Jason~F Shogren.
\newblock The benefits and costs of using social distancing to flatten the
  curve for covid-19.
\newblock {\em Journal of Benefit-Cost Analysis}, pages 1--27, 2020.

\bibitem{gros2020strategies}
Claudius Gros, Roser Valenti, Kilian Valenti, and Daniel Gros.
\newblock Strategies for controlling the medical and socio-economic costs of
  the corona pandemic.
\newblock {\em \href{Available as a working paper of the Clausen Center for
  International Business and
  Policy}{https://clausen.berkeley.edu/wp-content/uploads/2020/04/Corona.pdf}},
  2020.

\bibitem{shaw2020governance}
Rajib Shaw, Yong-kyun Kim, and Jinling Hua.
\newblock Governance, technology and citizen behavior in pandemic: Lessons from
  covid-19 in east asia.
\newblock {\em Progress in disaster science}, page 100090, 2020.

\bibitem{yamamoto2020apparent}
Naoki Yamamoto and Georg Bauer.
\newblock Apparent difference in fatalities between central europe and east
  asia due to sars-cov-2 and covid-19: Four hypotheses for possible
  explanation.
\newblock {\em Medical hypotheses}, 144:110160, 2020.

\bibitem{budish2020maximize}
Eric Budish.
\newblock Maximize utility subject to r<1: A simple price-theory approach to
  covid-19 lockdown and reopening policy.
\newblock Technical report, National Bureau of Economic Research, 2020.

\bibitem{goetheAnalyzer2020}
Claudius Gros, Fabian Schubert, and Carolin Roskothen.
\newblock \href{https://itp.uni-frankfurt.de/covid-19/}{Goethe Interactive
  Covid-19 Analyser}, 2020.

\bibitem{capasso1978generalization}
Vincenzo Capasso and Gabriella Serio.
\newblock A generalization of the kermack-mckendrick deterministic epidemic
  model.
\newblock {\em Mathematical Biosciences}, 42(1-2):43--61, 1978.

\bibitem{hethcote1991some}
Herbert~W Hethcote and P~Van~den Driessche.
\newblock Some epidemiological models with nonlinear incidence.
\newblock {\em Journal of Mathematical Biology}, 29(3):271--287, 1991.

\bibitem{github2020}
JHU-CSSE.
\newblock \href{https://github.com/CSSEGISandData/COVID-19}{John Hopkins Center
  of Systems Science and Engineering COVID-19 repository}, 2020.

\bibitem{gollier2020pandemic}
Christian Gollier.
\newblock Pandemic economics: optimal dynamic confinement under uncertainty and
  learning.
\newblock {\em The Geneva Risk and Insurance Review}, 45(2):80--93, 2020.

\bibitem{acemoglu2020multi}
Daron Acemoglu, Victor Chernozhukov, Iv{\'a}n Werning, and Michael~D Whinston.
\newblock A multi-risk sir model with optimally targeted lockdown.
\newblock Technical report, National Bureau of Economic Research, 2020.

\bibitem{rowthorn2020cost}
Robert Rowthorn and Jan Maciejowski.
\newblock A cost--benefit analysis of the covid-19 disease.
\newblock {\em Oxford Review of Economic Policy}, 36(Supplement\_1):S38--S55,
  2020.

\bibitem{strong2020estimation}
Aaron Strong and Jonathan~William Welburn.
\newblock {\em An Estimation of the Economic Costs of Social-Distancing
  Policies}.
\newblock RAND, 2020.

\bibitem{krueger2020macroeconomic}
Dirk Krueger, Harald Uhlig, and Taojun Xie.
\newblock Macroeconomic dynamics and reallocation in an epidemic.
\newblock {\em Covid Economics}, 5:21--55, 2020.

\bibitem{Eichenbaum2020}
Martin Eichenbaum, S{\'e}rgio Rebelo, and Mathias Trabandt.
\newblock The macroeconomics of epidemics.
\newblock {\em NBER Working Paper No. 26882}, 2020.

\bibitem{gros2020predicting}
Claudius Gros, Roser Valenti, Lukas Schneider, Benedikt Gutsche, and Dimitrije
  Markovic.
\newblock Predicting the cumulative medical load of covid-19 outbreaks after
  the peak in daily fatalities.
\newblock {\em medRxiv}, 2020.

\bibitem{atkeson2020will}
Andrew Atkeson.
\newblock What will be the economic impact of covid-19 in the us? rough
  estimates of disease scenarios.
\newblock {\em NBER Working Paper No. 26867}, 2020.

\bibitem{Brodeur}
Abel Brodeur, David Gray, Anik~Islam Suraiya, and Jabeen Bhuiyan.
\newblock \href{http://ftp.iza.org/dp13411.pdf}{A Literature Review of the
  Economics of COVID-19}.
\newblock {\em IZA DP No. 13411, IZA Institute of Labour Economics}, 2020.

\bibitem{baldwin2020keeping}
Richard Baldwin.
\newblock Keeping the lights on: Economic medicine for a medical shock.
\newblock {\em Macroeconomics}, 20:20, 2020.

\bibitem{coronaOutside}
Ivan Couronne.
\newblock
  \href{https://medicalxpress.com/news/2020-10-coronavirus-rare-impossible.html}{Catching
  coronavirus outside is rare but not impossible}.
\newblock {\em MedicalXpress}, 2020.

\bibitem{sajadi2020temperature}
Mohammad~M Sajadi, Parham Habibzadeh, Augustin Vintzileos, Shervin Shokouhi,
  Fernando Miralles-Wilhelm, and Anthony Amoroso.
\newblock Temperature and latitude analysis to predict potential spread and
  seasonality for covid-19.
\newblock {\em Available at SSRN 3550308}, 2020.

\bibitem{Holmstrom2020}
Princeton~University Bendheim Center~for Finance.
\newblock
  \href{https://bcf.princeton.edu/events/bengt-holmstrom-the-seasonality-of-covid-19/}{Bengt
  Holmstrom: The Seasonality of Covid-19}, 2020.

\bibitem{karapiperis2020assessment}
Christos Karapiperis, Panos Kouklis, Stelios Papastratos, Anastasia Chasapi,
  and Christos Ouzounis.
\newblock Assessment for the seasonality of covid-19 should focus on
  ultraviolet radiation and not `warmer days'.
\newblock 2020.

\bibitem{carlson2020misconceptions}
Colin~J Carlson, Ana~CR Gomez, Shweta Bansal, and Sadie~J Ryan.
\newblock Misconceptions about weather and seasonality must not misguide
  covid-19 response.
\newblock {\em Nature Communications}, 11(1):1--4, 2020.

\end{thebibliography}

\end{document}